\documentstyle{l-aa}
\input{epsf}

\begin{document}
  \thesaurus{02.          
              (02.01.4;   
               02.13.1;   
               08.14.1)   
            }
\title{Ionization of the hydrogen atom 
in strong magnetic fields.
}
\subtitle{
Beyond the adiabatic approximation
}

\author{A.Y.\,Potekhin$^{1,2,3}$
\and G.G.\,Pavlov$^{4,1}$
\and J.\,Ventura$^{2,3}$
}
\institute{
           $^1$ Ioffe Physical-Technical Institute,  
               194021, St.-Petersburg, Russia \\ 
           $^2$ Department of Physics, University of Crete, 
               710\,03 Heraklion, Crete, Greece \\
           $^3$ Institute of Electronic Structure and Laser, 
               FORTH, 711\,10 Heraklion, Crete, Greece \\
           $^4$ Pennsylvania State University, 525 Davey Laboratory,
                University Park, PA 16802, USA\\ 
}
\offprints{A.Y.\,Potekhin (Ioffe Institute)}

\date{Received 20 February 1996 / Accepted ~~ April 1996}

\maketitle

\newcommand{\vv}[1]{\mbox{\boldmath $#1$}}  
\newcommand{\ab}{a_{\rm B}}                 
\newcommand{\am}{a_{\rm m}}                 
\newcommand{\mel}{m_{\rm e}}                
\newcommand{\mpr}{m_{\rm p}}                
\newcommand{\req}[1]{(\ref{#1})}            
\newcommand{\dd}{{\rm d}}                   
\newcommand{\be}{\begin{equation}}
\newcommand{\ee}{\end{equation}}

\begin{abstract}
High magnetic fields in neutron stars, $B\sim 10^{11} - 10^{13}$~G,
substantially modify the  
properties of atoms and their interaction with radiation. 
In particular, the photoionization 
cross section becomes anisotropic and polarization dependent, being 
strongly reduced 
when the radiation is polarized perpendicular to the field. 
In a number of previous works based on the adiabatic 
approximation the conclusion was drawn that this transverse 
cross section vanishes 
for frequencies $\omega$ smaller than the electron cyclotron frequency
$\omega_c=eB/(\mel c)$. In other works 
(which employed a different form of the
interaction operator) appreciable finite values 
were obtained, $\sim \sigma_0 \gamma^{-1}$ near
the photoionization threshold, 
where $\sigma_0$ is the cross
section without magnetic field, and $\gamma = B/(2.35\times 10^{9}$~G). 
Since it is the transverse
cross section which determines the properties of radiation emitted
from neutron star atmospheres, 
an adequate interpretation of the neutron star 
thermal-like radiation requires a resolution of this controversy. 

In the present work 
we calculate the atomic wave functions for both discrete and continuum
states by solving the coupled channel equations 
allowing the admixture between different Landau levels, 
which provides much higher accuracy than the adiabatic approximation. 
This enables us to resolve the above  
contradiction in 
favour of the
finite transverse cross sections at $\omega <\omega_c$.
Moreover, for any form of the interaction operator
the non-adiabatic corrections appear to be substantial
for frequencies $\omega\ga 0.3\omega_c$. 
The non-adiabatic treatment of the continuum includes coupling between 
closed and open channels, 
which leads to the autoionization of quasi-bound energy levels
associated with the electron cyclotron
(Landau) excitations and gives rise to Beutler--Fano
resonances of the photoionization cross section. 
We calculate the autoionization widths of these quasi-bound
levels and compare them with the radiative widths. 
The correlation of the open channels is responsible for the 
modification of the cross section
above the Landau thresholds.
The results are important for investigations of the 
radiation emergent from the surface layers of neutron stars.


\end{abstract}

\markboth{Potekhin, Pavlov, \& Ventura: 
Ionization of H atom in strong magnetic fields}{}

\section{Introduction}                                    
\label{sect1}
The proper interpretation of the recently 
discovered surface radiation of isolated pulsars 
(e.g., \"Ogelman 1995) requires knowledge of the 
elementary processes in magnetic neutron star 
atmospheres (Pavlov et al.\ 1995). Among such processes, ionization 
of atomic hydrogen is important both conceptually, due to the 
simplicity of the hydrogen atom, and practically, because of the 
presumably strong gravitational stratification 
of the neutron star atmospheres. 

Although the photo- and autoionization 
of atoms have been 
thoroughly investigated at the magnetic field strengths 
$B\sim 10^5 - 10^9$~G (e.g., 
Merani et al.\ 1995, and references therein), none of these 
works can be directly extended to the typical pulsar field strengths 
$B \ga 10^{11}$~G. First, the atomic wave functions at the pulsar 
fields are nearly cylindrical, while at $B=10^9$~G the ground state 
atom is still nearly spherical. Second, the Landau levels at the 
pulsar fields are displaced much more distantly, which alters the 
spectrum qualitatively. These qualitative features arise when the 
parameter $\gamma = \hbar\omega_c/(2{\rm Ry}) = B/(2.35\times 
10^9~{\rm G})$ exceeds unity. Here $\omega_c = eB/(\mel c)$ is the 
electron cyclotron frequency, and ${\rm Ry} = \mel e^4/(2\hbar^2) = 
13.6$~eV is the Rydberg energy. 

The photoionization processes at $\gamma\gg 1$ have been considered 
in a number of papers: first analytically by Hasegawa \& Howard (1961) 
and Gnedin et al.\ (1974), and then both analytically and numerically 
by Schmitt et al.\ (1981), Wunner et al.\ (1983b), Mega et al.\ (1984), 
Miller \& Neuhauser (1991), 
Potekhin \& Pavlov (1993) (hereafter Paper~I), 
Bezchastnov \& Potekhin (1994) (B\&P), 
and Kopidakis et al.\ (1996) (KVH). 
All these papers, except for the two latter ones, 
employed the {\em adiabatic approximation\/} for the atomic 
wave functions. B\&P and KVH  studied the effect 
of atomic motion across the field 
on the photoionization process. 
Here we consider the particular case 
of atoms at rest. In this case the approach of KVH, 
who did not include electron cyclotron excitations, 
reduces back to the adiabatic one. 
There was no non-adiabatic treatment of the {\em final\/} state 
by B\&P or by KVH. 

The photoionization cross sections obtained in the cited papers 
appeared to be strikingly different. 
Some authors (e.g., KVH) concluded that, for the non-moving atom, 
the cross section for photons polarized 
perpendicular to the field 
vanishes in the most important frequency range $\omega < \omega_c$, 
while others (e.g., Paper~I) 
presented 
finite transverse cross sections which, 
although suppressed by the strong magnetic field, are still  
significantly larger than, e.g., the Thomson cross section.  Since 
the properties of radiation emitted from magnetic atmospheres are 
mainly determined by the transversely polarized photons (Pavlov et 
al.\ 1995), the different values of transverse cross sections should then 
result in quite different spectra and angular distributions of the 
radiation.

The first goal of the present paper is to resolve this contradiction,
which is of crucial importance for modeling the neutron star 
atmospheres. We shall prove that the discrepancy is caused solely by 
using the adiabatic approximation and is eliminated 
as soon as this approximation is abandoned. The zero values of the 
transverse cross section arise from using the so-called velocity 
form of the interaction potential, which (contrary to the 
alternative  length form) leads to inadequate results when used in 
the adiabatic approximation.  The two representations yield the same 
nonzero transverse cross sections when the more accurate  
non-adiabatic approach is applied. 

Our second principal goal is to
extend the consideration of the photoionization to the domains
of magnetic fields and photon energies 
where the adiabatic approximation does not provide
sufficient accuracy. For instance, more than a dozen 
radio pulsars
have magnetic fields in the range $\sim 10^{10}-10^{11}$~G;
the corresponding dimensionless fields, $\gamma \sim 4 - 40$,
although they exceed unity, are not sufficiently high
to neglect the non-adiabatic effects on the atomic properties.
The non-adiabatic effects are also expected to be considerable
for photon energies not very small as compared to the electron 
cyclotron energy, $\hbar\omega_c\simeq 1.2 (B/10^{11}~{\rm G})$~keV
(Paper I). The spectral flux is maximal at these X-ray energies
for effective temperatures $\sim 4\times 10^6 (B/10^{11}~{\rm G})$~K
characteristic of hot polar caps of radio pulsars. Such radiation
has been 
observed from a number of pulsars by
$ROSAT$, and new observations are under way with $ASCA$. For lower
temperatures, more typical at the surfaces of cooling neutron stars,
the radiation of such energies is still observable from nearby
objects. 

Among the non-adiabatic effects, particularly interesting are some 
new qualitative features of photoionization, which arise from 
including non-adiabatic terms in the final (continuum) state 
as well as in the initial state. In this approach, 
due to coupling to the continuum of lower Landau levels, 
the autoionization of quasi-bound atomic states,
enters naturally into the consideration, 
giving rise to the so-called Beutler--Fano 
resonances below the thresholds of the electron cyclotron 
excitations. Comparison
of the autoionization widths of the quasi-bound levels 
with the radiative widths shows that
autoionization is important at relatively low fields,
$B\la 3\times 10^{11}$~G. 
Our results demonstrate that the corresponding spectral features
are expected to be observable in the thermal-like spectra 
of neutron stars 
with the next generation of X-ray satellites (particularly,
$ASTRO$--$E$). 
In addition, above the electron cyclotron thresholds, 
different ionization channels are 
no longer independent of each other (as they were in the adiabatic  
approximation). The correlation of the channels 
is responsible for considerable modification of the
photoionization cross sections, which should be observable
in the spectra and light curves of the EUV/X-ray radiation
of neutron stars.

\section{Atomic wave functions}                        
\label{sect2}
\subsection{Coupled channel equations}
\label{sect2.1}
The early studies of the hydrogen atom 
in strong magnetic fields have 
been based on the adiabatic approximation 
(e.g., Canuto \& Ventura 1977). 
In this approximation the wave function
of the relative motion 
$\psi_{s\eta}(\vv{r})$
is factorized into a transverse part 
$\Phi_{ns}(\vv{r}_\perp)$
and a longitudinal part 
$g^{s\eta}_n(z)$, where $z$ is the relative 
coordinate along the magnetic field, and $\vv{r}_\perp = (x,y)$. 
The transverse part is just the Landau function which describes 
the transverse motion of a free electron in a magnetic field, 
$n$ being the Landau quantum number, and $s$ the negative 
of the $z$-projection of the electron orbital momentum ($n\geq 0$, 
$s\geq -n$). 
The quantum number 
$\eta=\pm 1$ refers to the $z$-parity of the wave function.
If the atom does not move as a whole across the field, which is the 
case of interest here, both $\eta$ and $s$ are exact quantum 
numbers, whereas $n$ can be considered as a ``good quantum number'' 
at $\gamma\gg 1$. The longitudinal wave function $g^{s\eta}_n(z)$ 
obeys a one-dimensional Schr\"odinger equation with a potential 
obtained by averaging the Coulomb potential over the transverse 
coordinates.
The total energy of an atom
is the sum of the longitudinal and transverse energies,
$E=E^\| + E_{ns}^\perp$,
with
\begin{equation}
   E^\perp_{ns} = \left[ n + (n+s)\mel/\mpr \right] \hbar\omega_c,
\label{transven}
\end{equation}
where the term $(n+s)\mel/\mpr$ accounts for the finite proton
mass (Herold et al.~1981).

In more recent studies the atomic wave functions have been calculated 
using various numerical approaches 
(e.g., Xi et al.\ 1992, and references therein). 
At $\gamma\ga 1$, the most convenient approach 
has been proposed by Simola \& Virtamo (1978) and developed 
by R\"osner et al.\ (1984) and Potekhin (1994) 
(Paper~II). It is based on 
the expansion of the wave function over the complete orthogonal 
set of the Landau functions $\Phi_{ns}$ in the $xy$ plane. 
Since, for non-moving atoms,  $s$ is 
an exact quantum number, the corresponding sum 
drops out, and we arrive at the reduced expansion 
\begin{equation}
         \psi_{s\eta}(\vv{r}) = \sum_{n'=n_{\rm min}}^\infty
 \Phi_{n's}(\vv{r}_\perp)\, g_{n'}^{s\eta}(z)~,
\label{expan}
\end{equation}
where $n_{\rm min}=0$ for $s\geq 0~$, and $n_{\rm min}=-s$ for $s<0~$.
The functions $g_n^{s\eta}(z)$ 
can be found from the following system of coupled
differential equations obtained by substituting 
the expansion \req{expan} into the Schr\"odinger equation, multiplying
it by $\Phi_{ns}^\ast(\vv{r}_\perp)$ and integrating over 
$\vv{r}_\perp$,
\begin{eqnarray}
&&
\left[-\frac{\hbar^2}{2\mu} \frac{\dd^2}{\dd z^2} + V_{nn}^{s}(z)
+E_{ns}^\perp - E\right]\, g_n^{s\eta}(z)
\nonumber\\
&&
=-\sum_{n'\neq n} V_{nn'}^{s}(z)\, g_{n'}^{s\eta}(z)~,
\label{coupledeq}
\end{eqnarray}
where $\mu$ is the reduced mass. The set of the effective
one-dimensional potentials is determined as
\begin{eqnarray}
&&\hspace{-.9em}
V_{nn'}^{s}(z) = \langle n s |-e^2/r|n' s\rangle_\perp
\equiv (\Phi_{ns} |-e^2/r | \Phi_{n's}) =
\nonumber\\
&&\hspace{-.9em}
   - {e^2\over\am\sqrt{2}} \int_0^\infty 
\hspace{-.5em}
   I_{n+s,n}(\xi) I_{n'+s,n'}(\xi) 
   (\xi+z^2/2\am^2)^{-1/2} \dd\xi~,
\label{effpot}
\end{eqnarray}
where $I_{nn'}$ are the Laguerre functions  (Sokolov \& Ternov 
1968), $\am=\ab\gamma^{-1/2}$ is the magnetic length,  and 
$\ab=\hbar^2/(\mel e ^2)$ is the Bohr radius.  Neglecting the 
non-diagonal ($n'\neq n$) effective potentials decouples the 
system and brings us back to the adiabatic 
approximation.  The choice of the maximum Landau number $n_{\rm max}$ 
for truncating expansion \req{expan} is dictated
by the desired accuracy --- the more terms one retains the 
higher accuracy of the wave function is provided. 

For any given set of the conserved quantities 
$s$ and $\eta$, each term 
in expansion \req{expan} is referred to as a 
{\em channel}, and the system \req{coupledeq} is the system of 
{\em coupled channel equations}.  The whole set of $n_{\rm 
max}-n_{\rm min}+1$ channels is separated into two groups. The 
first one includes {\em open channels} $n=n_{\rm min},\ldots 
,n_0-1$, for which $E >E_{ns}^\perp$. The second group embraces 
{\em closed channels} $n=n_0,\ldots , n_{\rm max}$, for which 
the opposite inequality, $E < E_{ns}^\perp$, holds.
\subsection{Bound states}
\label{sect2.2}
In the adiabatic approximation, the right hand side in \req{coupledeq}
vanishes and the channels are uncoupled. 
The bound states have negative longitudinal energies, i.~e., 
correspond to the closed channels. 
The discrete energy levels are enumerated at specified $s$, 
and $n$ by a longitudinal quantum number $\nu=0,1,2,\ldots$; 
the states having $\nu=0$ are known as 
tightly bound, while those with $\nu\geq 1$ are hydrogen-like. 
The longitudinal quantum number fully determines the $z$-parity of the 
state, $\eta=(-1)^\nu$, so that the quantum number $\eta$ becomes 
redundant and is usually omitted.

In the non-adiabatic approach,  
the RHS in \req{coupledeq} couples the state $|ns\nu\rangle$ 
into other channels, $n'\ne n$, 
comprising both bound states (closed channels) and continuum 
states (open). 
The wave functions 
$\psi_{ns\nu}(\vv{r})= \sum_{n'} \Phi_{n's}(\vv{r})_\perp\, 
g_{n'}^{ns\nu}(z)$ and the energy levels $E_{ns\nu}$ can be 
designated by a {\em leading} term $n'=n$ of the expansion 
\req{expan}, if $\gamma$ is large enough.  Only the states with 
$n=n_{\rm min}$ can be truly bound.  In other words, the energies 
$E_{ns\nu}$ correspond to truly bound states only when all the 
channels are closed. Coupling of the 
closed channels shifts the 
level energies from the adiabatic values and admixes 
higher Landau orbitals to the bound state.

A non-adiabatic computer code for calculating 
the bound state wave functions of the hydrogen atom moving in a 
strong magnetic field has been described in Paper~II. Here we apply this code 
(for the particular case of the non-moving atom) to obtain 
the initial state $|i\rangle$ of an atom subject to photoionization. 
\subsection{Continuum}                                 
\label{sect2.3}
The final atomic state $|f\rangle$ in the photoionization process 
lies in  the continuum, $E_f>E_{n_{\rm min}s}^\perp$,
i.~e., at least one associated 
channel is open. All previous publications 
on photoionization in very strong magnetic fields 
treat the final state  
adiabatically, i.e. $\langle \vv{r}|f\rangle = 
\Phi_{ns}(\vv{r}_\perp)\, g^{(f)}(z)$.
The simplest approach for the adiabatic continuum
wave function is the Born approximation, 
used by Gnedin et al.\ (1974) for the non-moving atom and generalized 
by KVH to the case of motion across the field. It assumes the 
longitudinal wave function of the final state 
to be of the form 
\begin{equation}
   g^{(f)}(z)=\exp(\pm{\rm i}k_nz), 
\label{born}
\end{equation}
where $k_n=\sqrt{2\mu (E_f-E_{ns}^\perp)}/\hbar$ is the electron 
wave number, $\pm$ denotes the direction of motion of the outgoing 
electron.  
A more accurate approach (e.~g., Paper~I) replaces 
$\exp(\pm{\rm i}k_nz)$ by a function 
$g^{(f)}(z)$ numerically determined from the corresponding
uncoupled equation; this function turns into \req{born} 
asymptotically, at $|z|\to\infty$  
(up to a logarithmic phase factor).  

In the non-adiabatic approach, the open (continuum) channel is 
coupled to closed channels and to other open channels 
(if the number of open channels $n_0 -n_{\rm min}> 1$).
This coupling of open and closed channels causes, in particular, 
autoionization of the quasi-bound states (states of positive
energy which would be bound in the absence of coupling --- see
Section 2.4). 
The coupling of different open channels can be conveniently treated
in terms of the {\em reactance matrix} $R_{nn'}$
(Seaton 1983). In the strong magnetic field case, 
the $R$-matrix can be introduced as follows.

System \req{coupledeq} of $~n_{\rm max}-n_{\rm min}+1~$ coupled
equations has 
(for a given parity $\eta$)  $~n_{\rm max}-n_{\rm min}+1~$
linearly independent solutions (basis vectors $g^{(n)s\eta}=
[g^{(n)s\eta}_{n_{\rm min}},\ldots , g^{(n)s\eta}_{n_{\rm max}}]$)
satisfying physically meaningful boundary
conditions. It is convenient to enumerate the solutions similarly
to the channels: $n=n_{\rm min}, ~n_{\rm min}+1,\ldots, n_{\rm max}$
(then $g_{n'}^{(n)s\eta}\to g_{n}^{(n)s\eta}\delta_{nn'}$ when the 
coupling switches off).  Each of these solutions forms a wave 
function $\psi_{s\eta}^{(n)}(\vv{r})= \sum_{n'} 
\Phi_{n's}(\vv{r}_\perp)\, g_{n'}^{(n)s\eta}(z)$ which 
comprises 
$~n_0-n_{\rm min}~$ open channels and $~n_{\rm max}-n_0+1~$ closed channels. 
For unbound solutions ($n=n_{\rm 
min},\ldots, n_0-1$) of a given $z$-parity, it is sufficient 
to construct {\em real\/} longitudinal wave functions
for positive $z$, which satisfy 
the following asymptotic conditions at $z\to +\infty$: 
\begin{equation}
   g^{(n,{\rm real})}_{n'} (z) \sim \delta_{nn'} \sin \phi_{n'}(z) 
   + R_{nn'} \cos\phi_{n'}(z) 
\label{g_asopen}
\end{equation}
         for $n'=n_{\rm min},\ldots,n_0-1$, 
         and 
         $g^{(n,{\rm real})}_{n'} (z) \to 0$ for 
         $n'=n_0,\ldots,n_{\rm max}$. 
(Hereafter, in this and the following subsections, we omit the  
indices $s$ and $\eta$ for brevity.)
In Eq.~\req{g_asopen}, 
\be
\phi_{n'}(z)=k_{n'} z + (k_{n'}\ab \mel/\mu)^{-1}\ln(k_{n'}z)
\label{phase}
\ee
is the $z$-dependent part of the phase.
With the continuum wave functions normalized to unity in
an interval of length $L_z$, 
the $R$-matrix satisfies the following symmetry relation,
\begin{equation}
   k_{n'} R_{nn'} = k_n R_{n'n}. 
\end{equation}
The real basis of the 
wave functions 
         $g^{(n,{\rm real})}_{n'} (z)$ 
is convenient for 
calculations, but it still is to be transformed into the basis 
of outgoing waves, appropriate to photoionization. 
This is done by analogy with the usual theory (Seaton 1983). 
For an electron outgoing in the positive $z$ direction, the following 
asymptotic condition at $z\to +\infty$ holds for the $n$-th solution: 
\begin{equation}
   g^{(n,{\rm out})}_{n'}(z) \sim 
   \delta_{nn'} \exp[\,{\rm i}\phi_{n'}(z)] 
   +S^\dagger_{nn'} \exp[-{\rm i}\phi_{n'}(z)],
\label{g_as_out}
\end{equation}
where $S^\dagger$ is the Hermitean conjugate scattering matrix. 
Now, if we compose a 
$(n_0-n_{\rm min}) \times (n_{\rm max}-n_{\rm min} +1)$ 
matrix function $G(z)$ 
of the elements $g^{(n)}_{n'}(z)$ (with 
$n$ the first and $n'$ the second subscript),
then we can obtain a set of the outgoing wave functions 
from the matrix equation
\begin{equation}
   G^{\rm (out)} = 2{\rm i} (1+{\rm i}R)^{-1} G^{\rm (real)}. 
\label{g_out}
\end{equation}
The $S^\dagger$ matrix is expressed in terms of the $R$ matrix as
  $ S^\dagger = -(1+{\rm i}R)^{-1} (1-{\rm i}R) $. 

Note that the longitudinal 
wave functions satisfying the asymptotic condition 
\req{g_as_out} should be multiplied by a common factor 
in order to ensure the correct normalization of the wave function 
$\psi^{(f,{\rm out)}}(\vv{r})$. It follows from the unitarity of the 
$S$-matrix that this factor equals $(2L_z)^{-1/2}$, 
where $L_z$ is the $z$-extension of the periodicity 
volume of the final state. 
\subsection{Autoionizing states}                      
\label{sect2.4}
The adiabatic and exact approaches treat 
in a fundamentally different way quasi-bound 
states associated with excited Landau levels. 
Since the adiabatic approximation allows no coupling between the 
Landau orbitals, a separate set of bound states appears below each 
Landau level $n$. 
These states, however, lie in the $n'<n$ continuum, and 
may therefore decay into the continuum via two processes. 
The first one is spontaneous emission of 
photons, which broadens these levels significantly in strong 
magnetic fields. This process has been 
thoroughly studied by Wunner et al.\ (1983a). The second one is 
{\em autoionization\/}, which could not be accounted for in the 
earlier work based on the adiabatic approach. 
In the non-adiabatic treatment, 
due to the coupling, the electron in a quasi-bound 
state can escape to infinity via any of the open channels $n'<n$. 

The quasi-bound autoionizing states manifest themselves as resonances
of the continuum wave function.
For weaker magnetic fields ($\gamma < 1$) such states 
have been studied, e.g., by Friedrich \& Chu (1983). 
Near a quasi-discrete level a resonance condition 
is satisfied, leading to a great amplification 
of the longitudinal coefficient $g_n(z)$. 
Thus, the shape of the wave function near the origin 
becomes similar to that calculated in the adiabatic approximation 
for the quasi-bound state. 
At large distances, where $g_n(z)$ decreases exponentially, 
the contribution of the orbitals $n'<n$ dominates, which 
can be interpreted as electron leakage from the quasi-bound state.

The general theory of autoionizing states has been described, e.~g.,
by Friedrich (1991). Here we briefly discuss it for the case
of the strongly magnetized hydrogen atom.
Let the electron energy $E$ be  close to an energy
$E_{n\nu}^{\rm ad}$ at which there would
be a bound state in the closed channel $n$ 
in the absence of channel coupling. 
To consider coupling of 
the channel $n$ to an open channel $n'<n$, 
one can retain the two corresponding terms in expansion \req{expan}
and two equations in the system \req{coupledeq}, assuming there are
no other quasi-discrete levels close to $E_{n\nu}^{\rm ad}$.
Let $g_{n'}^{\rm ad}(z)$ and $\bar{g}_{n'}^{\rm ad}(z)$ be the 
two linearly independent solutions for the uncoupled open
channel (e.~g., with asymptotic behaviour $g_{n'}^{ad}(z) \sim
\sin[\phi_{n'}(z)+\delta_{\rm ad}]$ and $\bar{g}_{n'}^{\rm ad}(z)
\sim \cos[\phi_{n'}(z) +\delta_{\rm ad}]$ at $z\to\infty$,
where $\delta_{\rm ad}$ is the adiabatic phase shift, and
$\phi_{n'}(z)$ is defined by Eq.~\req{phase}),
and $g_{n\nu}^{\rm ad}(z)$ be the solution for the uncoupled closed 
channel.  Then, by analogy with the nonmagnetic case, the solution 
of the two coupled channel equations can be presented in the form
\be
g_{n'}(z)=\cos\delta_c\,\, g_{n'}^{\rm ad}(z) + 
\sin\delta_c\,\, \bar{g}_{n'}^{\rm ad}(z)~,
\label{autoopen}
\ee
\be
g_{n\nu}(z)= 
 - \sin\delta_c\,\, 
\frac{U_{n\nu,n'}}{\Gamma_{n\nu,n'}^{\rm a}/2}\, g_{n\nu}^{\rm 
ad}(z)~, \label{autoclosed} \ee
where $\delta_c$ is an additional asymptotic phase shift due to
channel coupling,
\be
\tan\delta_c=-\frac{\Gamma_{n\nu,n'}^{\rm a}/2}{E-E_{n\nu}}~.
\label{couplingphase}
\ee
In Eqs.~\req{autoclosed} and \req{couplingphase}, 
\be
\Gamma_{n\nu,n'}^{\rm a}=
\frac{
2\mu
L_z}{\hbar^2k_{n'}}|U_{n\nu,n'}|^2
\label{autowidth}
\ee
is the (partial) {\em autoionization width} of the
quasi-bound state $|n\nu\rangle$,
\be
U_{n\nu,n'}=\int 
g_{n\nu}^{\rm ad}(z)\, V_{n,n'}(z)\, g_{n'}^{\rm ad}(z)\,
\dd z
\label{couplingmel}
\ee
is the coupling matrix element, and
$ E_{n\nu}$
is the resonance energy (slightly shifted from
$E_{n\nu}^{\rm ad}$ --- see Friedrich 1991).
Note that Eqs.~\req{autowidth} and \req{couplingmel} can be derived 
directly with the help of the usual perturbation theory.

Thus, it follows from Eqs.~\req{autoopen} and \req{autoclosed}
that coupling of the closed and open channels distorts
the open channel wave function and admixes the bound state
$|n\nu\rangle$ to the continuum. 
The strength of the admixture is given by the square of
the amplitude in front of the bound wave function $g_{n\nu}^{\rm 
ad}(z)$ in Eq.~\req{autoclosed}; its dependence on energy is 
determined by the resonance function (Breit--Wigner profile) which 
coincides with the derivative of the phase shift $\delta_c$ with 
respect to energy, 
\be
\frac{{\rm d}\delta_c }{{\rm d}E} = 
\frac{\Gamma_{n\nu,n'}^{\rm a}/2}{(E-E_{n\nu})^2 + 
(\Gamma_{n\nu,n'}^{\rm a}/2)^2}~.
\ee
The closed channel $n$ is most strongly coupled with 
the open channel $n'=n-1$. If we want to include coupling
with all the open channels, the total autoionization
width can be evaluated as $\Gamma_{n\nu}^{\rm a}=
\sum_{n'=n_{\rm min}}^{n-1}\Gamma_{n\nu,n'}^{\rm a}$.
As we shall see, the interference of the coupled states leads to 
the Beutler--Fano
resonances in the energy dependence of the radiative transitions.
%
\section{Interaction with radiation}                   
\label{sect3}
\subsection{Matrix element of the interaction}          
\label{sect3.1}
The cross section for ionizing an atomic state $|i\rangle$ 
into a continuum state $|f\rangle$ due to interaction 
with a photon 
of frequency $\omega$, wave vector $\vv{q}$, 
and polarization vector $\vv{e}$, can be written as 
(e.g., KVH)
\begin{equation}
   \sigma_{i\to f} = {3\over 8\alpha^3}\,{{\rm Ry}\over\hbar\omega}\,
   \sqrt{{\rm Ry}\over E_f^\|}\,{L_z\over\ab}\,
   |\langle f |\hat{M}|i\rangle|^2\,\sigma_{\rm Th}, 
\label{crsct}
\end{equation}
where 
$\alpha=e^2/(\hbar c)$, and $\hat{M}$ is the dimensionless interaction 
operator. Its ``velocity'' representation involves the kinetic 
momentum operator $\vv{\pi}=\vv{p}+
(e/2c)\vv{B}{\bf \times}\vv{r}$: 
\begin{equation}
   \hat{M}^{(\pi)} = \hat{M}^{(\pi)}_0 + \delta \hat{M}^{(\pi)}, 
\label{mpi_d}
\end{equation}
where
\begin{equation}
   \hat{M}^{(\pi)}_0 = \ab \exp({\rm i}\vv{q}\vv{r}) 
   \left[ {2\over\hbar}\,\vv{e}\cdot\vv{\pi} 
   - {\rm i}(\vv{q}\times\vv{e})_z \right],  
\label{mpi}
\end{equation} 
and $\delta \hat{M}^{(\pi)}$ denotes 
corrections of order 
$(\mel/\mpr)$. It will be shown that these corrections 
can be appreciable, when the velocity representation is used. 
Approximately one 
can write 
\begin{equation}
   \delta\hat{M}^{(\pi)} \approx {\mel\over\mpr}\ab
   {2\over\hbar}\,\vv{e}\cdot
   \left(\vv{p} - {e\over 2c} \vv{B}\times \vv{r} \right).
\end{equation}
The first term in the square brackets in Eq.~\req{mpi}
represents the electron current or ``velocity'' term in the interaction 
potential, while the second term 
corresponds to the interaction of the radiation magnetic field
with the magnetic spin moment of the electron.
We have neglected spin 
flip transitions, which are unimportant at $B\la 10^{13}$~G, 
according to 
Wunner et al.\ (1983b) and KVH, and fixed the electron spin
antiparallel to the magnetic field. Besides, we have 
omitted the term $\vv{e}\vv{q}$ (B\&P) from 
the square brackets, assuming the transverse polarization. 

Using Eq.\,\req{expan} for both atomic states, 
we obtain 
\begin{equation}
   \langle f|\hat{M}|i\rangle = \sum_{nn'} 
   \langle n,f| \langle n,s_f | \hat{M} | n',s_i \rangle_\perp 
   | n',i \rangle_\| ,
\label{m_expan}
\end{equation}
where 
the subscripts $\perp$ and $\|$ denote the transverse
(cf. Eq.~\req{effpot}) and longitudinal matrix elements,
respectively.
The inner (transverse) matrix element in Eq.\,\req{m_expan} 
can be calculated analytically, 
using the well known properties of the 
electron Landau quantum states (e.g., Canuto \& Ventura 1977), 
so that $\langle f|\hat{M}|i\rangle$ reduces to the sum 
of one-dimensional quadratures. 
\subsection{Transverse polarization: analytical 
consideration}
\label{sect3.2}
Using the commutation relations for the Hamiltonian, 
the matrix element $\langle f|\hat{M}|i\rangle$ can be rewritten 
in the ``length form'' (Paper~I): 
\begin{eqnarray}
   \langle f | \hat{M}^{(\pi)} | i \rangle &=& 
   \langle f | \hat{M}^{(r)} | i \rangle, 
\label{mpi_mr}
\\
   \hat{M}^{(r)} &=& \hat{M}^{(r)}_0 + \delta \hat{M}^{(r)}, 
\\
   \hat{M}^{(r)}_0 &=& 
   {\rm i} \ab \exp({\rm i}\vv{q}\vv{r}) 
   \left[\rule{0mm}{3ex}\right.
   {2\mel\omega\over\hbar}\,\vv{e}\cdot\vv{r}
   \left(\rule{0mm}{3ex}\right.
   1-{\hbar\omega\over 2\mel c^2}  
\nonumber\\[-.3ex] &
   -  & 
   {\vv{q}\cdot\vv{\pi}\over \mel\omega}
   \left.\rule{0mm}{3ex}\right)
   -(\vv{q}\times\vv{e})_z
   \left.\rule{0mm}{3ex}\right].
\label{mr}
\end{eqnarray}
Here, as well as in Eq.\,\req{mpi_d}, $\delta \hat{M}$ 
denotes corrections $\sim \mel/\mpr$. 

For exact atomic states $| i \rangle$ and $| f \rangle$, 
the two representations 
are equivalent. 
However, the equivalence \req{mpi_mr} breaks down in the adiabatic 
approximation. 
The most striking difference occurs for transitions 
within the ground Landau state when the photons are polarized 
perpendicular to the magnetic field. It 
was recently confirmed by KVH that in this case 
$\langle 0,s_f | \hat{M}^{(\pi)} | 0,s_i \rangle_\perp = 0$ 
identically, if one neglects the small corrections due to 
$\delta \hat{M}^{(\pi)}$. 
One actually finds an exact cancellation of the contributions arising 
from the velocity and the spin interaction terms in Eq.~\req{mpi}. 
At the same time, the representation \req{mr} 
leads to
a non-zero cross section\footnote{ 
The spin term $(\vv{q}\times\vv{e})_z$, 
omitted in Paper~I, cannot be responsible for this result, 
as was assumed by KVH. 
Indeed, 
the transverse cross section in Paper~I remains non-zero 
even for photons directed along the field, 
in which case the spin term turns 
out to vanish identically.}. 
Some analytical estimates help to resolve this 
apparent contradiction 
(see also Appendix A of Paper I). 

Since series \req{m_expan} converges rapidly at 
$\gamma\gg 1$, we can expect that it is sufficient 
to keep only the leading terms in it. 
In the velocity representation, the zero-order term 
$\langle 0,s_f | \hat{M}^{(\pi)}_0 | 0,s_i \rangle_\perp$ 
vanishes (KVH). 
Retaining the first-order terms, we obtain 
\begin{eqnarray}
   \langle f | \hat{M}^{(\pi)} | i \rangle &\approx & 
   \langle 1,f |\, \langle 1,s_f | \hat{M}^{(\pi)}_0 
   | 0,s_i \rangle_\perp\, |0,i\rangle_\| 
\nonumber\\ &
   +  &    
   \langle 0,f | \, \langle 0,s_f | \hat{M}^{(\pi)}_0
   | 1,s_i \rangle_\perp \, |1,i\rangle_\|
\nonumber\\ &
   +   &    
   \langle 0,f | \, \langle 0,s_f | \delta \hat{M}^{(\pi)} 
   | 0,s_i \rangle_\perp \, |0,i\rangle_\|.
\label{mpi_appr}
\end{eqnarray}
Which of the three terms 
in Eq.\,\req{mpi_appr} 
dominates, depends on the magnetic field 
strength and 
transition considered. 
 
In the length representation, the zero-order term dominates:
\begin{equation}
   \langle f| \hat{M}^{(r)} | i \rangle \approx 
   \langle 0,f| \, \langle 0,s_f | \hat{M}^{(r)}_0 | 0,s_i 
   \rangle_\perp \, |0,i \rangle_\|. 
\label{mr_appr}
\end{equation}
Let us consider, for simplicity, the dipole approximation, 
$q \to 0$.
It has been shown in Paper I (see also Sect.~4) that this approximation
can be safely used at $\hbar\omega
\ll \alpha \mel c^2 \ln\gamma \sim 300\ln\gamma$~Ry.
For $\theta=0$, Eq.~\req{mpi_appr} then yields 
\begin{eqnarray}
   \langle f | \hat{M}^{(\pi)} | i \rangle &\approx& 
   2{\rm i}\,\sqrt{\gamma} \left[ \rule{0mm}{2.3ex} \right. 
   e_+ (\langle 0,f | 1,i \rangle_\| 
   + \zeta\,\sqrt{s_f}) 
   \delta_{s_f, s_i+1}
\nonumber\\ & 
- 
 & 
   e_- (\langle 1,f | 0,i \rangle_\|
    + \zeta\,\sqrt{s_i})  
   \delta_{s_f,s_i-1} \left. \rule{0mm}{2.3ex}\right], 
\label{mpi1}
\end{eqnarray}
where $e_\pm = (e_x \pm {\rm i} e_y) /\sqrt{2}$ 
are the cyclic components of the vector $\vv{e}$, 
and 
\begin{equation}
   \zeta = {\mel\over\mpr}\langle 0,f|0,i\rangle_\|.
\label{zeta}
\end{equation}
The terms with $\zeta$ represent 
the leading contribution due to $\delta \hat{M}^{(\pi)}$. 
Analogously, from Eq.\,\req{mr_appr} we obtain 
\begin{eqnarray}
   \langle f | \hat{M}^{(r)} | i \rangle & \approx & 
   2{\rm i}\,\sqrt{\gamma} \,{\omega\over\omega_c} 
   \left[\rule{0mm}{2.3ex}\right.
   e_+\,\sqrt{s_f}\, \delta_{s_f,s_i+1} 
\nonumber\\ &
+
 &
   e_- \,\sqrt{s_i}\,  \delta_{s_f, s_i-1} 
   \left.\rule{0mm}{2.3ex}\right]
   \langle 0,f | 0,i \rangle_\| .
\label{mr0}
\end{eqnarray}
Conditions under which Eqs.~\req{mpi1} and \req{mr0} 
give similar cross sections are obtained by equating their 
right-hand sides: 
\be
   \langle 0,f | 1,i \rangle_\| \approx 
  \sqrt{s_f} \left( {\omega\over\omega_c} - {\mel\over\mpr} \right)
   \langle 0,f | 0,i \rangle_\|, 
\label{0_1} 
\ee
or 
\be
   \langle 1,f | 0,i \rangle_\| \approx 
   - \sqrt{s_i} \left( {\omega\over\omega_c} + {\mel\over\mpr} \right) 
   \langle 0,f | 0,i \rangle_\|, 
\label{1_0}
\ee
where $s_f = s_i + 1$ or $s_i = s_f + 1$, respectively. 
The term $\mel/\mpr$ comes from 
$\delta \hat{M}^{(\pi)}$ through Eqs.\,\req{mpi1}, \req{zeta}. 
Note that
the terms $\mel/\mpr$ 
can give a substantial correction (especially
near the photoionization thresholds, $\hbar\omega \sim
\ln^2\gamma$~Ry) in very strong fields, when $\gamma \ln^{-2}\gamma$
is not negligible in comparison with $\mpr/\mel$.
 
Our analytical estimates (Appendix~A) and numerical 
results (Sect.~4) confirm that the approximate 
relationships 
(\ref{0_1}), (\ref{1_0}) are indeed 
satisfied if $\omega\ll\omega_c$. 
Thus, in the adiabatic approximation 
(for $n=n'=0$), the length 
representation of the interaction matrix element, employed 
in Paper~I, enables one to calculate the leading contribution 
to the transverse cross section 
which is missed when the velocity representation is used. 
On the other hand, it 
makes no difference which representation 
is used if the post-adiabatic corrections are included.
\subsection{Interference of open and closed channels: Beutler--Fano
resonances}
As we have discussed in Sect.~2.4, coupling of a closed channel $n$
to open channels, particularly to the channel $n-1$, results
in a resonance admixture of the quasi-bound state to the continuum.
Interference of the two final coupled states leads to
{\em Beutler--Fano resonances} in the photoabsorption spectrum
(e.~g., Friedrich 1991). According to Eqs.~\req{autoopen}
and \req{autoclosed}
                (for $n'=n-1$), 
the transition matrix 
element, for the final electron
energy close to the energy of the quasi-bound state,  can be
presented as
\be
\langle f|\hat{M} | i\rangle = 
M_1\, \cos\delta_c\, (1-q\, \tan\delta_c)~,
\ee
where 
\be
q=\frac{M_2}{M_1}
\frac{U_f}{\Gamma_f^{\rm a}/2}
\ee
is the  shape parameter of the Beutler--Fano resonance
($U_f\equiv U_{ns_f\nu_f,n-1}$, $\Gamma_f^{\rm a}\equiv 
\Gamma_{ns_f\nu_f,n-1}^{\rm a}$ ---
see Eqs.~\req{autowidth} and \req{couplingmel}),
\be
M_1=
(g_{n-1,s_f}^{\rm ad}|\langle
n-1,s_f |\hat{M}|
n_is_i\rangle_\perp |g_{n_is_i\nu_i})_\|
\ee
is the matrix element in the absence of coupling,
\begin{eqnarray}
M_2 &=& 
- (\Gamma_f^{\rm a}/2U_f)
(\bar{g}_{n-1,s_f}^{\rm ad}|\langle
n-1,s_f|\hat{M}|
n_is_i\rangle_\perp |g_{n_is_i\nu_i})_\| 
\nonumber\\
&+&
(g_{ns_f\nu_f}^{\rm ad}|\langle
ns_f
|\hat{M}|
n_is_i\rangle_\perp |g_{n_is_i\nu_i})_\|
\end{eqnarray}
is the coupling correction.
The cross section is proportional to
\begin{eqnarray}
|\langle f|\hat{M}| i\rangle |^2 
&=&
 |M_1|^2\, 
\cos^2\delta_c\,\, |1-q\, \tan\delta_c|^2
\nonumber\\
&=&
 |M_1|^2\, F(\epsilon, q)~,
\end{eqnarray}
where
\be
F(\epsilon, q)=\frac{|q+\epsilon|^2}{1+\epsilon^2}
\ee
is the Beutler--Fano function,
\be
\epsilon=-\cot\delta_c = 2(E-E_{ns_f\nu_f})/\Gamma_f^{\rm a} 
\ee
is the reduced energy. 
The function $F(\epsilon, q)$ tends to unity in the 
very far wings of the resonance, at $|\epsilon|
\gg {\rm max}(1, |q|)$. If $q$ is real (e.~g.,
in the dipole approximation), it  turns to zero
at $\epsilon=-q$ and is maximal ($F_{\rm max}=1+q^2$) at 
$\epsilon=1/q$.
In the limit of very weak coupling, $U_f\to 0$, we have $q\propto 
U_f^{-1} \to \infty$, $q^2\Gamma_f^{\rm a}$ remains finite, and the 
Beutler--Fano resonance turns into the delta-function (superimposed 
onto the bound-free continuum) which describes the bound-bound 
transition in the absence of coupling.

\section{Results}
We present here 
example photoionization cross sections, calculated 
in accordance with the approach described above. 
In all cases the hydrogen atom is assumed 
to be initially in its ground state. 
Basically we have employed the velocity 
representation of the interaction matrix element, including into 
$\hat{M}^{(\pi)}$ all the non-dipole and finite-mass corrections 
(Paper~I), as well as the spin term (KVH). The results are compared 
with those obtained with the length representation, the adiabatic, and 
dipole approximations. Some details of the numerical techniques used
are given in Appendix B. 

Our first principal result is that at $\omega\ll\omega_c$ 
Eqs.\,\req{0_1} and \req{1_0} do hold with a 
high accuracy 
(the difference does not exceed 
1--2\% 
at 
 $\hbar\omega\la 30$ Ry), 
thus confirming the considerations of Sect.\,3.2 and Paper~I. 
In the calculations we have also included higher terms 
of the expansions \req{expan} and \req{m_expan}; however, 
the channels $n,n'> n_f+1$ 
proved to be unimportant. 
We have also checked the 
importance of the spin term, 
and found it to be negligible, 
by switching it on and off numerically.
On the contrary, the corrections due to the finite proton mass 
are appreciable. If the terms $\mel/\mpr$ on the right-hand side 
of Eqs.\,\req{0_1}, \req{1_0} were omitted, then the
inaccuracy would increase to 10\% at $\gamma=1000$. 
Clearly, this would introduce an inaccuracy of about 20\% 
in the cross section. 

\begin{figure}[t]
\begin{center}
\leavevmode
\epsfysize=95mm 
\epsfbox{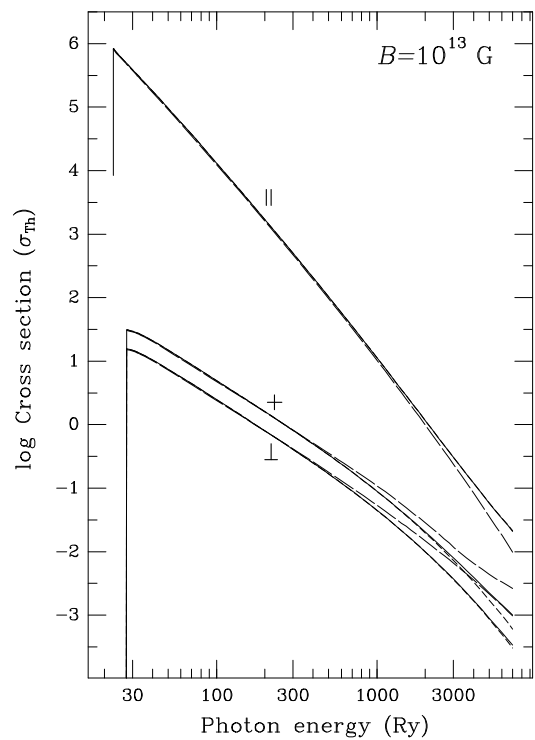}
\end{center}
\caption[ ]{
Total cross sections of the photoionization of the ground state 
H atom at the magnetic field strength 
$B=10^{13}$~G. 
The curves are labelled by the symbols $+$, $\perp$, and $\|$, 
corresponding respectively to $\sigma_+$ (right circular polarization 
at $\theta=0^{\mbox{\tiny 0}}$), $\sigma_\perp$ (polarization vector 
transverse to $\vv{B}$ at $\theta=90^{\mbox{\tiny 0}}$), 
and $\sigma_\|$ (polarization along $\vv{B}$ 
at $\theta=90^{\mbox{\tiny 0}}$). 
Numerical results (solid lines) are compared 
with the dipole approximation (short-dashed lines) 
and with the adiabatic results of Paper~I (long-dashed lines)
}
     \label{fig1}
\end{figure}

Figure~\ref{fig1} shows the cross section as a function of the photon 
energy $\hbar\omega$ for photoionization from
the ground level, at a magnetic field strength  
$B=10^{13}$~G 
($\hbar\omega_c = 8509$~Ry);
the cross sections plotted are
for photons propagating along the magnetic field
($\theta=0^\circ$) with the right 
circular polarization 
($\sigma_+$), and propagating perpendicular to the field
($\theta=90^\circ$)
with the linear polarizations 
parallel ($\sigma_\|$) and perpendicular ($\sigma_\perp$) 
to $\vv{B}$.
Short-dashed lines correspond to the dipole approximation. 
We see that at $\theta=90^{\mbox{\tiny 0}}$ 
the non-dipole corrections are unimportant 
for $\sigma_\|$ and $\sigma_\perp$ at $\hbar\omega\la 10^4$~Ry. 
This confirms the validity of the transverse dipole approximation 
in this energy range and  
justifies negleting the spin term, 
as it depends on the transverse wave vector only.
 On the other hand, the 
inaccuracy of the longitudinal dipole approximation becomes 
perceptible at 
$\hbar\omega\ga (2-3)\times 10^3$~Ry, 
which is seen from the deviation 
of the short-dashed line from the solid one for $\sigma_+$.

For comparison, the results of Paper~I
(length representation) which do not 
include the non-adiabatic and spin terms are shown by 
long-dashed lines. 
The agreement with the 
adiabatic approach of Paper~I,
which involved the length form of
the interaction matrix element, is fairly good 
at $\hbar\omega\la 10^3$~Ry.
Note that the velocity form would lead 
to vanishing $\sigma_+$ and $\sigma_\perp$ (KVH). 

\begin{figure*}[t]
\begin{center}
\leavevmode
\epsfysize=95mm 
\epsfbox{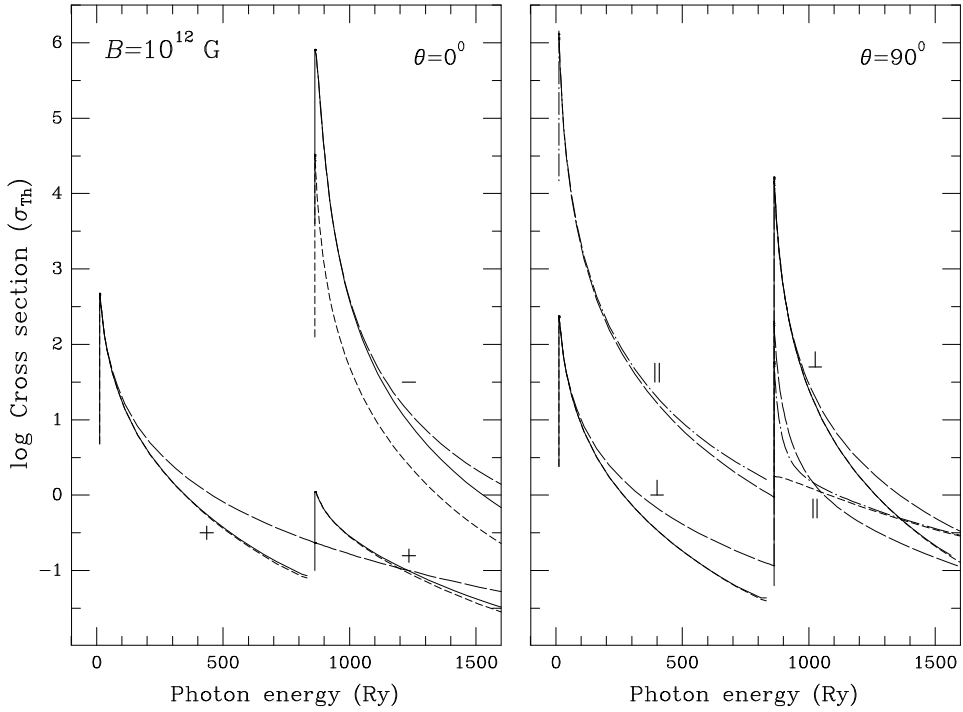}
\end{center}
\caption[ ]{
Same as Fig.~\ref{fig1} for $B=10^{12}$~G. 
The symbol `$-$'  
corresponds to the left circular polarization 
at $\theta=0^{\mbox{\tiny 0}}$.  
Numerical results for $\sigma_-$ and $\sigma_\|$ are 
plotted with the dot-dashed line, 
numerical results for $\sigma_+$ and $\sigma_\perp$ 
with the solid line. The long-dashed and short-dashed lines 
correspond to the adiabatic (Paper~I) and dipole 
approximations, respectively. 
{\bf a} Circular polarization ($\sigma_\pm$) 
at $\theta=0^{\mbox{\tiny 0}}$. 
{\bf b} Linear polarization ($\sigma_\perp,~\sigma_\|$) 
at $\theta=90^{\mbox{\tiny 0}}$
}
\label{fig2}
\end{figure*}

The agreement with the adiabatic results becomes worse 
as $\omega$ approaches $\omega_c$ from below, 
which is caused by the growing role of the ``side'' terms 
(closed channels $n>0$) in Eq.\,\req{m_expan}. 
Figure~\ref{fig2} demonstrates the cross sections at $B=10^{12}$~G
($\hbar\omega_c=850.87$~Ry).
We see that the adiabatic approximation (long dash) 
may serve only as an order-of-magnitude estimate at
$\omega\sim\omega_c$,
whereas at $\omega
\la (0.2-0.3)\omega_c$ the agreement is good again. 
For the magnetic field chosen, 
the channel $n=1$ opens
at the (threshold) energies $\hbar\omega=862.79$, 863.25 and 863.72 
Ry for $s_f=-1$, 0 and 1, respectively (these would be the $n=1$ 
thresholds for the left, right and longitudinal polarizations in the 
dipole approximation).  Immediately above the thresholds, the 
adiabatic approximation for $\sigma_-$ and $\sigma_\perp$ becomes 
sufficiently accurate again, as has been predicted
in Paper I.  Photoionization from
the ground state is  strictly forbidden for the left polarization
below the $n=1$ threshold.
Above the threshold, 
the corresponding cross section 
$\sigma_-$ appears to be strongly underestimated 
in the dipole approximation (see the left panel of Fig.~\ref{fig2}). 
The importance of the factor 
$\exp({\rm i}\vv{q}\vv{r})$ in this case is explained by 
approximate coincidence of the photon and electron wave numbers, 
as discussed in Paper~I. 
The dipole approximation is inadequate also for $\sigma_\|$ 
(right panel of Fig.~\ref{fig2}) 
above the $n=1$ threshold
because it misses the channel $n=1,~s_f=-1$, 
which gives the main contribution just 
at these energies.
For this polarization, the adiabatic approximation 
(long-dashed lines) proves to be inaccurate as well, mainly 
because of the admixture of the (open) channels $n=0$ and $n=1$.

\begin{figure*}[t]
\begin{center}
\leavevmode
\epsfysize=95mm 
\epsfbox{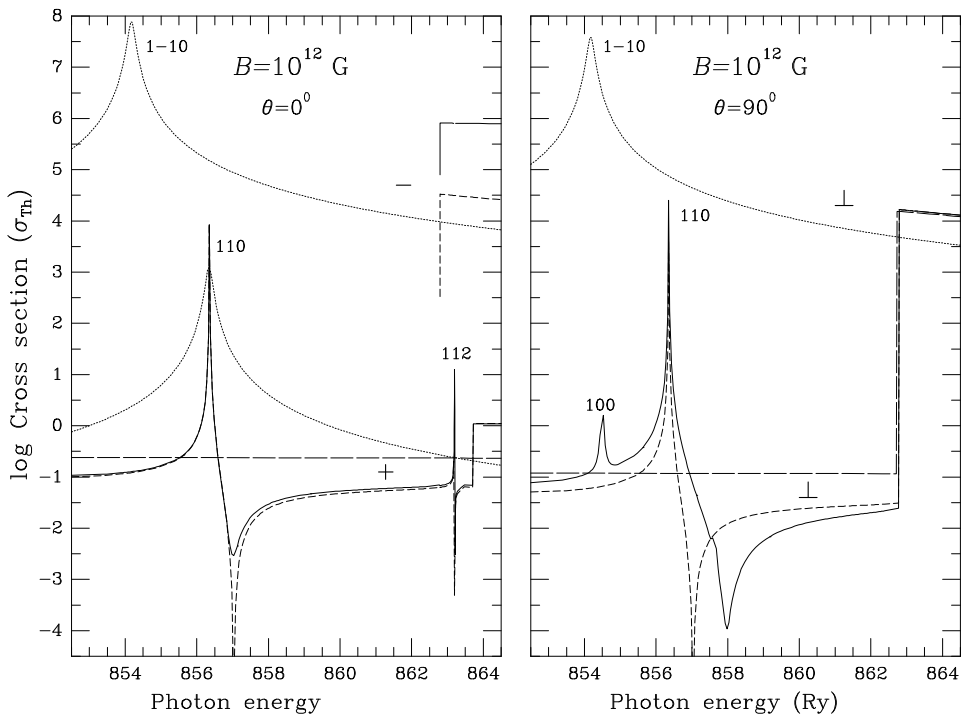}
\end{center}
\caption[ ]{
Same as Fig.~\ref{fig2} in the energy range near the Landau threshold. 
{\bf a} Circular polarization ($\sigma_\pm$) 
at $\theta=0^{\mbox{\tiny 0}}$. 
{\bf b} Linear transverse polarization ($\sigma_\perp$) 
at $\theta=90^{\mbox{\tiny 0}}$. 
Superimposed on the same diagram, 
the dotted curves show the ``bound-bound'' absorption
profiles calculated without allowance for coupling of the closed and
open channels; the shapes of these profiles are determined by the
radiative deexcitation of the upper levels (see text for details)
}
\label{fig3}
\end{figure*}

The solid and short-dashed curves for the non-adiabatic cross sections 
presented in Fig.~\ref{fig2} 
do not include values within $\sim 30$~Ry just below the $n=1$ 
threshold.  Within this gap complex resonance structures
arise in the photoionization cross sections, 
too narrow to be resolved 
in the scale of Fig.~\ref{fig2}. An example of such structures 
is shown in an expanded scale in Fig.~\ref{fig3}.
The resonances are associated with the quasi-bound autoionizing
states $|1,s_f,\nu_f\rangle$ admixed to the continuum 
$|0,s_f\rangle$.  Their shape  in the dipole approximation
(short-dashed lines)
is typical for the Beutler--Fano resonances (see Sect.~3.3)
with a negative $q$ parameter: 
the peaks (at $E=E_{1s_f\nu_f}+
(2q)^{-1}\Gamma_{1s_f\nu_f}^{\rm a}$) 
are followed by troughs
(at $E=E_{1s_f\nu_f}-q \Gamma^{\rm a}_{1s_f\nu_f}/2$). If
$|q|\gg 1$ for a given resonance
(which is fulfilled for the 
resonances shown in Fig.~\ref{fig3}), then
the FWHM of the peak coincides with the autoionization
width $\Gamma_{1s_f\nu_f}^{\rm a}$ (see Sect.~2.4),
and the distance between the peak maximum and 
the trough minimum is $\simeq |q|\Gamma_{1s_f\nu_f}^{\rm a}/2\gg 
\Gamma_{1s_f\nu_f}^{\rm a}$.
For instance, for the main peak of $\sigma_+(E)$, which is
associated with the autoionizing state $|1,1,0\rangle$,
we have $\Gamma_{110}^{\rm a}\simeq 0.01$~Ry and $q\simeq -200$.
Resonances of the same nature have been obtained previously for 
the case of lower magnetic fields, $\gamma\ll 1$ (cf. Delande
et al.~1991; O'Mahoni \& Mota-Furtado 1991). For  strong
magnetic fields (and, consequently, higher photon energies),
non-dipole corrections change the shape of the resonances
so that the shape becomes different for different
polarizations and angles $\theta$.
In particular, the parameter $q$ is no longer real,
which results in  non-zero values of the trough minima
shifted from their positions obtained in the dipole
approximation. 
An additional effect entering when we go beyond the dipole approximation
is the appearance of additional dipole-forbidden 
peaks overlapping with the dipole-allowed resonances. For instance,
for $\sigma_\perp$, 
the peak at $\hbar\omega\simeq 856.4$~Ry, associated with the 
quasi-bound state $|1,1,0\rangle$, is preceded by another one, 
at $\hbar\omega\simeq 854.5$~Ry,
attributable to the state $|1,0,0\rangle$, transitions to which are 
dipole-forbidden (due to a ``transverse dipole'' selection rule, 
see also Ventura et al.\ 1992).
Similarly, dipole-forbidden transitions 
(i.e. opening of the $n=1, s=1$ continuum) are responsible
for the jump of $\sigma_+$ at 863.7~Ry.

 The second, much weaker and narrower resonance 
(its autoionization width is only $\simeq 1.5\times 10^{-4}$~Ry) 
in Fig.\ 3a is associated with the 
hydrogen-like quasi-bound state $|1,1,2\rangle$. 
It is not observed in Fig.\ 3b 
because of the orders of magnitude stronger background 
absorption due to the transition to the
continuum state $n_f=1$, $s_f=-1$, 
which is allowed at $\theta\neq 0$ and whose threshold 
lies below this hydrogen-like quasi-bound level. In fact, there 
exist other Beutler--Fano resonances, related to the transitions
to more excited quasi-bound states, but they are much too weak and 
narrow to play any role in the computed spectrum, 
so that we do not display them here.  For the same reason, we 
also do not show the autoionizing resonances of $\sigma_\|$.  
For the longitudinal polarization, only transitions to 
the odd states are allowed at $\theta=90^{\mbox{\tiny 0}}$. 
Therefore only hydrogen-like autoionizing states can contribute to 
$\sigma_\|$. The corresponding resonances, however, are extremely 
weak and narrow ($\Gamma^{\rm a} < 10^{-4}$~Ry).

Figure 4 shows the cross sections at a weaker field, 
$B = 10^{11}$~G. This field strength was not considered 
in Paper~I because the adiabatic approximation 
may become too crude in this case. The present 
non-adiabatic treatment allows us to include this
(and lower) field strength(s) into 
the consideration. Several Landau thresholds appear 
in the observationally relevant energy range in this case. 
Figure~5 demonstrates the resonances 
associated with these thresholds.  
\begin{figure*}[t]
\begin{center}
\leavevmode
\epsfysize=95mm 
\epsfbox{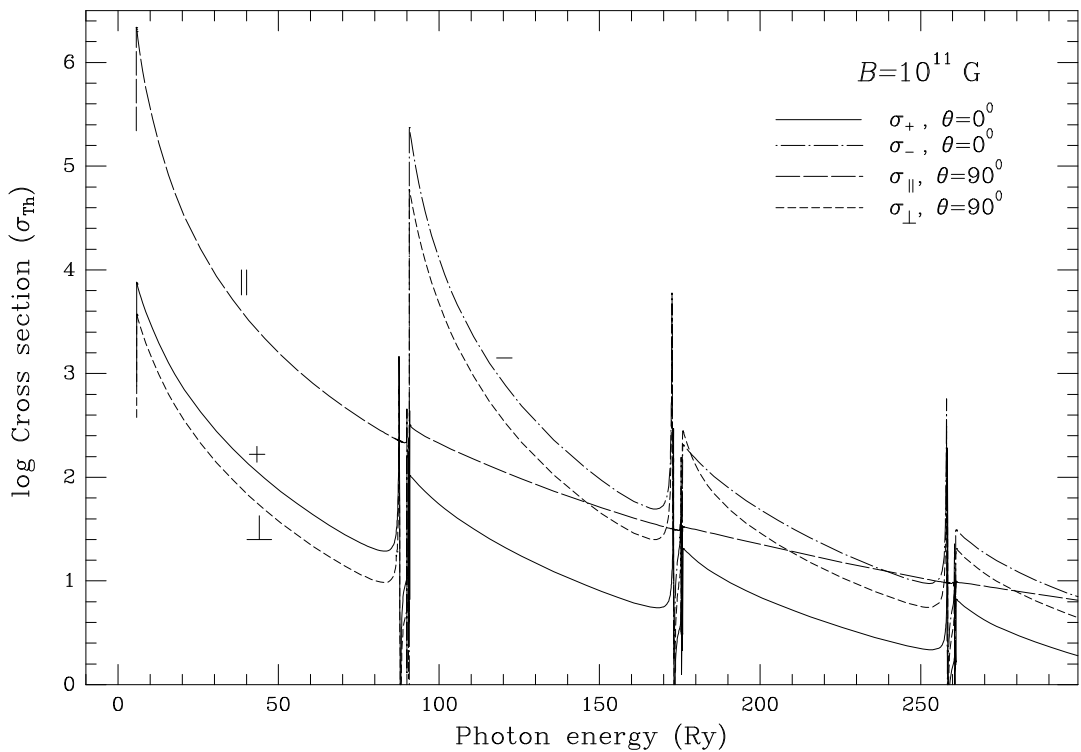}
\end{center}
\caption[ ]{
Total cross sections of the photoionization of the ground state 
H atom at the magnetic field strength 
$B=10^{11}$~G. Solid and dash-dot lines correspond
to the right and left circular polarizations, respectively,
at the incidence angle $\theta=0^{\mbox{\tiny 0}}$; short-dash
and long-dash lines correspond to the  
transverse and longitudinal linear polarizations 
at $\theta=90^{\mbox{\tiny 0}}$
}
\label{fig4}
\end{figure*}

The cross section for the right circular polarization 
at $\theta=0^{\mbox{\tiny 0}}$ 
(solid lines in Figs.\,4, 5) reveals a relatively broad resonance
below each $n$-th Landau threshold
associated with the tightly bound autoionizing states 
$|n,1,0\rangle$ 
(their peaks lie at $\hbar\omega=$ 87.6, 173.0 and 258.4 Ry,
and the autoionizing widths are $\Gamma^{\rm a}=$ 
0.014,
0.016, and 0.014 
Ry,
for $n=1,2,3$, respectively). 
A sequence of weaker and narrower resonances is further seen to 
converge to a corresponding threshold. These are 
related to the even 
hydrogen-like autoionizing states $|n,1,2\rangle$,
$~|n,1,4\rangle$, $\ldots$

Analogous features for the left polarization (dot-dashed curves),
associated with the states $|n,-1,0\rangle$, $|n,-1, 2\rangle$, 
$\ldots$, are seen for $n\geq 2$. The states $|1,-1,\nu\rangle$ are 
not coupled to the  continuum (if the motion across the magnetic 
field is neglected) and do not contribute to the photoionization 
cross section. On the contrary, coupling of the states 
$|n,-1,0\rangle$ for $n\geq 2$ is relatively strong (e.~g., 
$\Gamma_{n-10}^{\rm a}= 0.014$ 
and 0.016 Ry for $n=2$ and 3,
respectively), and the corresponding resonances dominate in 
Fig.~5b,c.   The cross section $\sigma_\perp$ 
for the transverse polarization at $\theta=90^\circ$ (short dashes)
shows autoionization resonances associated with both
$|n,1,\nu\rangle$ (for $n\geq 1$) and $|n,-1,\nu\rangle$
(for $n\geq 2$) because $e_\perp =
         (e_++e_-)/\sqrt{2}$ 
is 
composed of both circular polarizations.
\begin{figure*}
\begin{center}
\leavevmode
\epsfysize=95mm 
\epsfbox{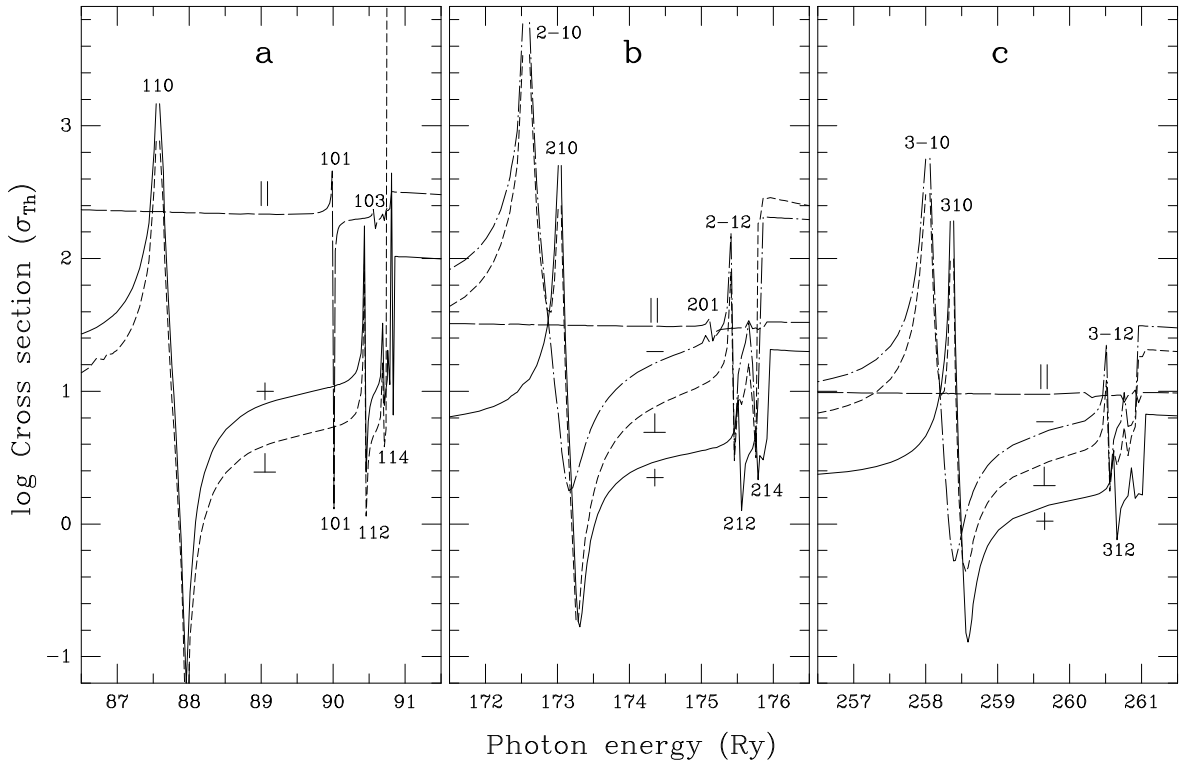}
\end{center}
\caption[ ]{
Same as Fig.~4 for the energy ranges near the first three 
Landau thresholds (panels {\bf a}, {\bf b} and {\bf c}, 
respectively)
}
\label{fig5}
\end{figure*}

For the longitudinal polarization (long dashes), 
transitions to odd states are only allowed 
at $\theta=90^{\mbox{\tiny 0}}$. 
The strongest resonance (at $\hbar\omega=90.0$ Ry) 
in Fig.~5a corresponds to the state $|1,0,1\rangle$
($\Gamma_{101}^{\rm a}\simeq 0.0008$ Ry). 
Other odd-state resonances (barely seen in Fig.\ 5a) 
are considerably weaker. 
At higher Landau thresholds, all resonances of this type 
are weak and narrow. 

The Beutler--Fano resonances in Figs.~3 and 5 were calculated
assuming that autoionization is the only channel for decay
of the quasi-bound states, so that other mechanisms which could
lead to additional broadening of the resonances can be neglected.
However, an excited state can always decay via spontaneous
emission of a photon, and it is known (e.~g., Wunner et al.~1983a)
that the rate of the radiative decay can be very high
in strong magnetic fields. 
The relative importance of the radiative
and autoionization decays is determined by the relation between
the autoionization width $\Gamma^{\rm a}$ and radiative width 
$\Gamma^{\rm r}$. If $\Gamma^{\rm a}\gg \Gamma^{\rm r}$, then
most electrons excited to the quasi-bound state 
         rapidly escape into 
the continuum, and the shape of the photoabsorption
resonances is determined by the autoionization.
In the opposite case,
spontaneous emission occurs faster than autoionization,
so that absorption of radiation at resonance energies is not
accompanied by photoionization, and the shape of the resonance
is described by the Lorentz profile of the width $\Gamma^{\rm r}$.
Thus, it is important
to compare $\Gamma^{\rm a}$ and $\Gamma^{\rm r}$ for a given
level in order to evaluate which of the two   
 processes is more important.

The (total) autoionization widths for a few levels
$|1,s,\nu\rangle$ and $|2,s,\nu\rangle$ are plotted as a function of 
$B$ in Fig.~6.  We see that they decrease with increasing $B$, being 
quite different for different autoionizing states. To understand 
qualitatively the behaviour 
of the widths, consider, for instance, the dependence of
the coupling matrix element $U_{1s\nu,0}$, Eq.~\req{couplingmel},
 which couples the $n=1$ quasi-bound
states to the $n=0$ continuum,
on the dimensionless magnetic field $\gamma$.
A characteristic length, $\sim a_M=\ab\gamma^{-1/2}$,
of the effective potential $V_{10}(z)$ in Eq.~\req{couplingmel},
which determines the limits of integration over $z$, is
much smaller than a typical size, $\sim a_\|~$ 
($\sim \ab/\ln\gamma$ for $\nu=0$,  and $\sim \ab(\nu+1)/2$
for $\nu>0$),  of
the bound wave function $g_{1\nu}^{\rm ad}(z)$. Therefore,
 in the integrand of Eq.~\req{couplingmel}, we have
 $g_{1\nu}^{\rm ad}(z)\sim a_\|^{-1/2}$ for the even 
quasi-bound states,
and $g_{1\nu}^{\rm ad}(z)\sim a_\|^{-1/2} (a_M/a_\|)$ for the
odd states.
The continuum wave function $g_{0}^{\rm ad}(z)$
depends on the product $kz \sim (2\gamma)^{1/2} (z/\ab)\la 
\sqrt{2}$.
Taking into account that a characteristic magnitude
of $V_{10}(z)$ is $\sim e^2/a_M\sim (e^2/\ab)\gamma^{1/2}$,
we arrive at the following estimates for the autoionization
widths at $\gamma\gg 1$:
$\Gamma_{1s\nu}^{\rm a}\propto \gamma^{-1/2}\ln\gamma$ for the
tightly bound states, $\propto \gamma^{-1/2}$ for the hydrogen-like
even states, and $\propto\gamma^{-3/2}$ for the hydrogen-like
odd states. 
Similar behaviour is observed for the autoionization widths of
the states with $n>1$.
%
\begin{figure*}[t]
\begin{center}
\leavevmode
\epsfysize=78mm 
\epsfbox{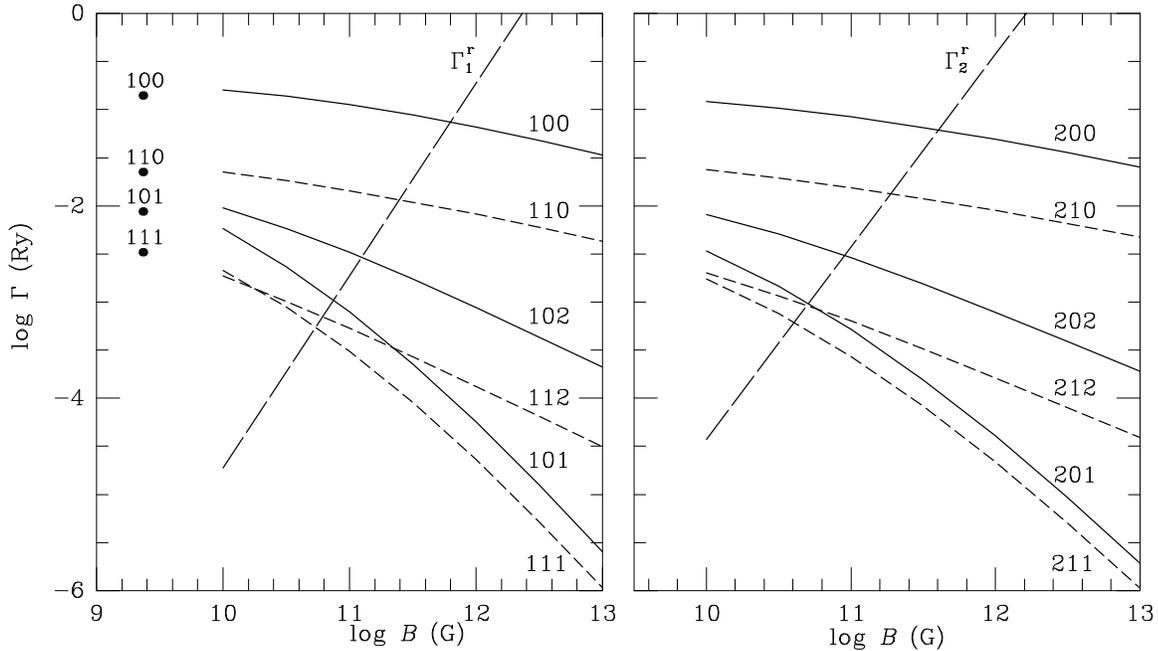}
\end{center}
\caption[ ]{
Autoionization widths of a few states $|n,s,\nu\rangle$ (labels near
the curves) vs.~magnetic field. For $n=2$, the total widths,
$\Gamma_{2s\nu}^{\rm a}=\Gamma_{2s\nu,0}^{\rm 
a}+\Gamma_{2s\nu,1}^{\rm a}$ are shown. The autoionization widths of 
the states $|2,-1,\nu\rangle$ coincide with those of the states 
$|1,1,\nu\rangle$.  The long-dashed curves show the radiative widths 
 $\Gamma_n^{\rm r}$ of the levels with $n=1$ and 2. The dots in the 
left panel show the values of the autoionization widths of the states 
(from top to bottom) $|1,0,0\rangle$, $|1,1,0\rangle$, 
$|1,0,1\rangle$ and $|1,1,1\rangle$ calculated by Friedrich and Chu 
(1983) for $B=2.35\times 10^9$~G ($\gamma=1$)
}
\label{fig6}
\end{figure*}

The radiative widths are mainly determined
by the transitions $|ns\nu\rangle \to |n-1,s+1,\nu\rangle$;
at $\gamma\gg 1$ the widths depend only on the Landau quantum number 
$n$ and the magnetic field,
\be
\Gamma_n^{\rm r}=\frac{8}{3} \alpha^3 n \gamma^2~,
\label{gammarad}
\ee
where $\alpha$ is the fine structure constant. In fact, \req{gammarad}
coincides with the well-know cyclotron width for cyclotron transitions
of free electrons (e.~g., Daugherty \& Ventura 1978). 
We see from Fig.~6
 that the radiative width exceeds all the autoionization widths
at $B\ga 6\times 10^{11}$ G for $n=1$, and $B\ga 4\times 10^{11}$~G 
for $n=2$. This means that at very high magnetic fields the 
quasi-bound states are destroyed via radiative decay rather than via 
autoionization, and the shape of the resonances of the 
photoabsorption cross section is determined predominantly by the 
 radiative broadening. At lower magnetic field, the shape of the 
resonances may be determined by different mechanisms for different 
quasi-bound levels. For instance, $\Gamma^{\rm a}\gg \Gamma^{\rm r}$
for the leading resonances in Fig.~5 (e.g. associated with the
$|1,1,0\rangle$, $|2,-1,0\rangle$, $|2,1,0\rangle$ states), whereas
weaker resonances are subject to stronger radiative broadening.

To illustrate the expected effect of radiative de-excitation on
the photoabsorption  spectra, we added to Fig.~3 Lorentz
profiles of the ``bound-bound'' transitions to
the states
$|1,1,0\rangle$ and $|1,-1,0\rangle$ (dotted lines). The height and
the shape of the profiles are determined by the radiative
width ($\Gamma_1^{\rm r}=0.19$~Ry) and the oscillator strengths
($f_{000,110}=3.1\times 10^{-5}$, $f_{000,1-10}=1.987$).
Note that $|1,-1,0\rangle$ 
is the truly bound state (for atoms at rest), and 
transitions to it are actually the strongest amongst
all the transitions from the ground state, so that
this ``cyclotron absorption'' dominates in the photoabsorption
spectrum near the $n=1$ Landau threshold. It should 
also be mentioned that the transition  
$|0,0,0\rangle \to |1,1,0\rangle$
is forbidden in the adiabatic approximation; its 
oscillator strength is provided by the admixture of
$|1,0,0\rangle$ to the ground state,
and $|0,1,0\rangle$ to the excited state.
\section{Conclusions}
We have studied photoionization cross sections of the 
hydrogen atom in magnetic fields $B\sim 10^{11}-10^{13}$~G, 
typical for 
pulsars. 
We have used 
exact interaction matrix elements, including effects of finite 
proton mass, non-dipole and spin interaction terms. 
Unlike previous authors, we use {\em non-adiabatic} wave functions 
for the initial and final states of the atom.
This accurate treatment yields the following conclusions.

First, it resolves the 
acute contradiction of previous works 
concerning the cross section for photons polarized transversely 
to the magnetic field at energies 
$\hbar\omega < \hbar\omega_c$. 
This cross section is finite,
being orders of magnitude larger than $\sigma_{\rm Th}$
near the threshold, 
in agreement with Paper~I. This is shown 
to have no connection with neglecting the spin interaction, 
as was assumed by KVH.  
Moreover, at $\omega\ll\omega_c$ the present results 
         nearly 
coincide quantitatively with those in Paper~I. 
We have proven that the zero values of
$\sigma_+$ and $\sigma_\perp$, 
obtained by a number of authors following Schmitt et al.\ (1981), 
do not represent the reality 
but arise from their using the {\em velocity\/} representation 
of the interaction matrix elements in combination with 
the {\em adiabatic\/} approximation for the wave functions. 
This combination led those authors to miss 
the main contribution in the cross section.

Second, the non-adiabatic treatment includes 
autoionization of the quasi-discrete levels
associated with the Landau excitations.
These levels are considered as truly discrete in the adiabatic 
approximation. 
Autoionization manifests itself in the photoionization
cross sections as Beutler--Fano resonances 
near the quasi-discrete levels;
the shape of the resonances is determined
by correlation between closed and open channels.
We have calculated the
autoionization widths for the most important resonances
in a wide range of  magnetic fields
and have shown that they exceed
the radiative widths if the magnetic field is
not too strong, $B\la 10^{11}$ G.

Third, we have shown that the adiabatic results significantly 
deviate from the exact ones not only near the resonances, 
but in a rather wide range of photon energies, 
unless the 
condition $\omega\ll\omega_c$ is satisfied. 
Above the Landau threshold,
this deviation is due to correlation between 
different open photoionization 
channels.

In this paper we restrict ourselves to the particular case 
of an atom which does not move across the field. 
It has been shown by Pavlov \& M\'esz\'aros (1993),
B\&P and KVH that transverse atomic motion leads to the opening of 
additional channels of ionization. In the particular case of 
{\it slow\/}  motion, however, our present consideration
allows one to decide between the small 
transverse cross sections presented by 
KVH and orders of magnitude 
larger cross sections obtained by B\&P --- 
in favour of the larger ones. 
\begin{acknowledgements}
This work was supported in part by INTAS (Grant 94-3834). 
A.Y.P.\ acknowledges partial support from 
RBRF (Grant 96-02-16870) 
and hospitality of the University of Crete and the 
Research Center of Crete (FORTH). 
The work of G.G.P.\ was supported by NASA Grant NAG5-2807. 
J.V.\ wishes to acknowledge partial support under NATO 
Grant Nr.\,CRG.931446. 
\end{acknowledgements}

\appendix
\addtocounter{section}{1}
\section*{Appendix A: 
 derivation of Eqs.~(\ref{0_1}), (\ref{1_0})}    
In this section, we derive an approximate relation 
between non-adiabatic corrections which, according 
to Eq.~(\ref{mpi1}), determine 
the interaction matrix element in the velocity representation 
for the transverse polarization, 
and the adiabatic overlap integral 
$\langle 0,f | 0,i \rangle_\|$ which enters Eq.~(\ref{mr0}) 
for the matrix element in the length representation. 

Let us consider the quantum states $|a\rangle$ and $|b\rangle$ 
($n_a=n_b=0$) belonging to adjacent $s$-manifolds: 
$s_a=s$, $s_b=s+1$. The longitudinal 
coefficients in Eq.\,\req{expan} are given by the set of 
coupled differential equations \req{coupledeq} with the
effective potential \req{effpot}.
At $\gamma\gg 1$, the non-diagonal 
effective potentials are small compared to the diagonal ones, 
and the non-adiabatic admixtures $g_{n\neq 0}$ are small 
respectively. Retaining only the leading terms in the equation 
for $g_0(z)$, we have 
\begin{eqnarray}
&&\hspace{-.9em}
 \left[ {\hbar^2\over 2\mu}\,{\dd^2\over\dd z^2} 
   +E_a - s{\mel\over\mpr}\hbar\omega_c \right] 
   g_0^{(a)}(z) 
   =
   V_{00}^{(s)}(z) g_0^{(a)}(z), 
\label{a0}\\
&&\hspace{-.9em}
   \left[ {\hbar^2\over 2\mu}\,{\dd^2\over\dd z^2} 
   +E_b - (s+1){\mel\over\mpr}\hbar\omega_c \right] 
   g_0^{(b)}(z)
\nonumber\\ 
&&\quad
   =
   V_{00}^{(s+1)}(z) g_0^{(b)}(z). 
\label{b0}
\end{eqnarray}
Keeping the first-order terms in the equation for 
$g_1^{(a)}(z)$, we obtain 
\begin{eqnarray}
&&\hspace{-.9em}
  \left[ {\hbar^2\over 2\mu}\,{\dd^2\over\dd z^2} 
   +E_a - 
   \left(1+s{\mel\over\mpr}\right)\hbar\omega_c \right]
   g_1^{(a)}(z) 
\nonumber\\
&&
   =
   V_{11}^{(s)}(z) g_1^{(a)}(z) + V_{01}^{(s)}(z) g_0^{(a)}(z).
\label{a1}
\end{eqnarray}
Here the terms $V_{n1}g_n$ with $n\geq 2$ are neglected. 

Multiplying Eq.~\req{a0} by $g_0^{(b)}(z)$ and 
Eq.~\req{b0} by $g_0^{(a)}(z)$, subtracting them by terms 
and integrating over $z$, we arrive at the approximate relation 
\begin{eqnarray}
&&\hspace{-.9em}
  \left(E_a - E_b + {\mel\over\mpr}\hbar\omega_c\right)
   \langle 0,a | 0,b \rangle_\| 
\nonumber\\
&&
 = 
   \langle 0,a | V_{00}^{(s)} - V_{00}^{(s+1)} | 0,b \rangle_\|. 
\label{ab0}
\end{eqnarray}
Analogously, from Eqs.~\req{b0} and \req{a1} we obtain 
\begin{eqnarray}
&&\hspace{-.9em}
   \left[E_a - E_b - 
   \left(1 - {\mel\over\mpr}\right) \hbar\omega_c\right]
   \langle 1,a | 0,b \rangle_\| \nonumber\\
&&
=
   \langle 0,a | V_{01}^{(s)} | 0,b \rangle_\| 
  +
   \langle 1,a | V_{11}^{(s)} - V_{00}^{(s+1)} | 0,b \rangle_\|.
\label{ab1}
\end{eqnarray}
The second term on the right-hand side should be omitted 
in the approximation considered. Indeed, according to 
Eq.~\req{effpot}, the difference of two diagonal 
effective potentials can be expressed 
in terms of non-diagonal ones,
\begin{eqnarray}
   V_{11}^{(s)}(z) - V_{00}^{(s+1)}(z) &=& (s+1)^{-1/2}
\nonumber\\
&\times&
   \left[ \sqrt{s+2}\,V_{02}^{(s)}(z) - V_{01}^{(s)}(z) \right], 
\end{eqnarray}
and thus the last term in Eq.~\req{ab1} 
represents a second order correction.  
Analogously, the difference on the right-hand 
side of Eq.~\req{ab0} equals 
\begin{equation}
   V_{00}^{(s)}(z) - V_{00}^{(s+1)}(z) = 
   (s+1)^{-1/2} V_{01}^{(s)}(z).
\end{equation}
Comparison of Eqs.~\req{ab0} and \req{ab1} yields now: 
\begin{equation}
   {\langle 1,a | 0,b \rangle_\| \over 
   \langle 0,a | 0,b \rangle_\| }
   \approx
   \sqrt{s+1}\,
   {E_a - E_b + (\mel/\mpr)\hbar\omega_c 
   \over
   E_a - E_b - (1-\mel/\mpr)\hbar\omega_c }. 
\end{equation}
Applying this result to radiative transitions between a lower 
state $|i\rangle$ and an upper state $|f\rangle$, 
neglecting $\mel/\mpr$ in the denominator, and assuming 
$\omega\ll\omega_c$, we arrive at Eqs.~\req{0_1} and 
\req{1_0}. 
\addtocounter{section}{1}
\setcounter{equation}{0}
\section*{Appendix B: details of computation}    
\subsection{Solving the Schr\"odinger equation}   
\label{sect_A1}
Bound state wave functions are calculated using the 
multiconfigurational Hartree--Fock technique described 
in Paper~II. 

Continuum wave functions are sought in the form \req{expan}, 
separately for each $z$-parity. 
The set of Eqs.\,\req{coupledeq} for the longitudinal coefficients 
is rearranged in two coupled subsystems, 
which are solved by two-step 
 iterations. 
In the first step, 
the equations for the open channels ($n,n'=n_{\rm min},\ldots,n_0-1$) 
are solved, the contribution to the right-hand side from 
the other group of orbitals ($n'=n_0,\ldots,n_{\rm max}$) being fixed. 
In the second step, 
the longitudinal wave functions of the 
closed channels ($n,n'\geq n_0$) are adjusted to the 
open-channel functions ($n'<n_0$) found in the preceding step. 
The procedure is then repeated; 
typically, a few such iterations are sufficient 
to reach convergence. 
The only exception occurs in the resonance energy region, 
 where the number of iterations increases up to 20--30, 
 and the method finally 
fails in narrow energy regions corresponding 
to the very top of the peaks. 
Nevertheless, 
as can be seen in our figures, we are still able to trace 
    substantial parts of the resonance profiles.

The first step of each iteration is performed by the outward 
integration, employing the explicit fourth-order Runge--Kutta 
scheme (e.g., Fletcher 1988) for the vector function 
$\vv{g} = (g_0^{(f)},\ldots,g_{n_0-1}^{(f)})$. 
The integration extends to the point 
$z_0 \sim 10^2\am$, 
where the off-diagonal effective potentials become negligible. 
The second step, solving the longitudinal equations for 
$g_{n\geq n_0}$ provided that $g_{n<n_0}$ are fixed, 
does not differ from that described in Paper~II. 
After the iterative process ends, one extra integration
for each orbital is 
required to proceed beyond $z_0$, where the orbitals 
are already uncoupled. 
\subsection{Rearrangement}                     
The outward integration for the open channels is performed 
with the initial conditions arbitrarily chosen as 
$g_n(0) = g'_n(0) = 0$ for all $n<n_0$ except  
$n=j$, whereas we chose 
$g_j(0)=1$ or $g'_j(0)=1$ 
depending on parity. In this way we obtain an arbitrary 
set of $n_0$ linearly independent wave functions 
$\psi^{(j,{\rm arb})}(\vv{r})$, $j=n_{\rm min},\ldots,n_0-1$. 
This set has to be rearranged, in order to meet 
the asymptotic conditions \req{g_asopen}: 
\begin{equation}
   g^{(n,{\rm real})}_{n'}(z) = \sum_{j=n_{\rm min}}^{n_0-1} 
   a_{nj} g^{(j,{\rm arb})}_{n'}(z), 
\label{rearrange}
\end{equation}
where the coefficients $a_{nj}$ 
         constitute 
the rearrangement matrix $A$. 
In order to obtain $A$, first, the integration is extended  
to a point $z_{\rm as} \gg e^2/\min(E_f - E^\perp_{ns_f})$, 
where the asymptotic behaviour is reached. The minimum longitudinal energy 
here is to be taken over all open channels, i.e. for 
$n=0,\ldots,n_0-1$. An outermost part 
of this integration may be performed by a faster scheme, 
cf.\ Paper~I. 
Then, using Eq.\,\req{rearrange} together with 
Eq.\,\req{g_asopen}, we arrive at the algebraic system: 
\begin{eqnarray}
&&\hspace{-.9em}
   \sum_{j=n_{\rm min}}^{n_0-1} g^{(j,{\rm arb})}_{n'}(z_{\rm as}) 
   a_{nj} - \cos\phi_{n'}(z_{\rm as}) R_{nn'} 
\nonumber\\[-1ex] && \qquad
   = \delta_{nn'} \sin\phi_n(z_{\rm as}), 
\nonumber\\ &&\hspace{-.9em}
   \sum_{j=n_{\rm min}}^{n_0-1} {\dd\over\dd z}
   g^{(j,{\rm arb})}_{n'}(z_{\rm as})
   a_{nj} + \phi'_{n'}(z_{\rm as}) \sin\phi_{n'}(z_{\rm as})R_{nn'}  
 \nonumber\\[-.5ex] && 
   = \delta_{nn'} \phi'_n(z_{\rm as})\cos\phi_n(z_{\rm as}), 
   ~~ n, n'=0,\ldots,n_0-1,
\label{systema}
\end{eqnarray}
the phase $\phi_n$ being defined in Eq.\,\req{phase}. 
For each given $n$, the $2(n_0-n_{\rm min})\times 2(n_0-n_{\rm 
min})$ system \req{systema} is solved to get the $n$th row of the 
matrices $A$ and $R$. Since the matrix of this algebraic system does 
not depend on $n$, the FACT/SOLVE code (Fletcher 1988) is most 
useful. 

Longitudinal matrix elements, which enter Eq.\,\req{m_expan}, 
are calculated along with the functions $g^{(j,{\rm arb})}$ 
at the last stage of the integration 
for each specific $j$. 
They still need to be transformed into the matrix elements 
for the outgoing states. According to 
Eqs.\,\req{rearrange} and \req{g_out}, this transformation of the 
array of matrix elements is performed 
by the matrix 
${\rm i}\sqrt{2/L_z}(1+{\rm i}R)^{-1}A$, which acts on the subscript 
$j$ related to a channel. The square root multiplier ensures the 
necessary normalization (Sect. 2.2).


%

\end{document}